\documentclass[aps,preprint,amsmath,amssymb]{revtex4}

\usepackage{graphicx}
\begin{document}

\title{Charged current reactions in the NuSOnG and a test of neutrino-W couplings}

\author{A. B. Balantekin}
\email[]{baha@physics.wisc.edu} \affiliation{Department of Physics,
University of Wisconsin, Madison, Wisconsin 53706, USA}

\author{\.{I}. \c{S}ahin}
\email[]{isahin@wisc.edu} \email[]{isahin@science.ankara.edu.tr}
\affiliation{Department of Physics, University of Wisconsin,
Madison, Wisconsin 53706, USA}
 \affiliation{Department of
Physics, Faculty of Sciences, Ankara University, 06100 Tandogan,
Ankara, Turkey}

\begin{abstract}
We explore the physics potential of NuSOnG experiment to probe new
physics contributions to $W\mu\nu_\mu$ couplings in inverse muon
decay and charged current deep inelastic scattering processes. We
show that NuSOnG provides two order of magnitude improvement in
$W\mu\nu_\mu$ compared to LEP bound at $95\%$ C.L.. Combining the
results presented here with our earlier results, we obtain a
significant improvement in the limits on neutrino flavor
universality violation in neutrino-W and neutrino-Z couplings.
\end{abstract}

\maketitle

\section{Introduction}
The physics potential of the recently proposed NuSOnG (Neutrino
Scattering on Glass) experiment \cite{Adams:2008cm}
is under intensive study. It is suggested that NuSOnG
can look beyond the standard model at terascale energies by making
precision electroweak and QCD measurements. NuSOnG can place
stringent constraints on possible non-standard neutrino
interactions. Potential of the NuSOnG experiment to probe new physics
contributions to $Z\nu\nu$ couplings in muon-neutrino electron
elastic and neutral-current deep-inelastic scattering processes have
been studied in Ref. \cite{Balantekin:2008rc}. It was shown that NuSOnG
experiment has significant potential to probe neutrino-Z couplings. In
the current paper we investigate NuSOnG sensitivity limits on non-standard
neutrino-W couplings via inverse muon decay and charged current deep
inelastic scattering.

 Non-standard $W\ell \nu$ couplings are
constrained approximately by up to an order of $10^{-2}$ from the
LEP data via W boson decay to leptons \cite{Yao:2006px}. Precision
measurement of muon decay also constrains the coupling with about
the same order \cite{muondecay}. There is an extensive literature on
non-standard interactions of neutrinos
\cite{Davidson:2003ha,Bell:2005kz,Mohapatra:2005wg,Mohapatra:2006gs,de
Gouvea:2007xp,Perez:2008ha}.  Our
formalism is based on an effective lagrangian approach presented in
Ref. \cite{Buchmuller:1985jz}. We do not {\it a priori} assume the
universality
of the coupling of neutrinos to W. Combining our results with the
results obtained in Ref. \cite{Balantekin:2008rc} we discuss the
potential of NuSOnG experiment to probe possible universality
violation at neutrino-Z and neutrino-W couplings.

In the effective lagrangian approach of
Ref.\cite{Buchmuller:1985jz}, possible deviations from the standard
model that may violate flavor universality of the neutrino-V (V=Z,W)
couplings are described by the following $SU(2)_L\otimes U(1)_Y$
invariant dimension-6 effective operators:
\begin{eqnarray}
\label{eop1}
O_j=i(\phi^\dagger D_\mu \phi)(\bar \psi_j \gamma^\mu \psi_j)\\
O^\prime_j=i(\phi^\dagger D_\mu \vec{\tau} \phi)\cdot(\bar \psi_j
\gamma^\mu \vec{\tau} \psi_j) \label{eop2}
\end{eqnarray}
where $\psi_j$ is the left-handed lepton doublet for flavor
$j=e,\mu$ or $\tau$; $\phi$ is the scalar doublet; and $D_\mu$ is
the covariant derivative, defined by

\begin{eqnarray}
D_\mu=\partial_\mu+i\,\frac{g}{2}\,
\vec{\tau}\cdot\vec{W}_\mu+i\,\frac{g^\prime}{2}\,YB_\mu .
\end{eqnarray}
Here $g$ and $g^\prime$ are the $SU(2)_L$ and $U(1)_Y$ gauge
couplings, $Y$ is the hypercharge and the gauge fields $W^{(i)}_\mu$
and $B_\mu$ sit in the $SU(2)_L$ triplet and $U(1)_Y$ singlet
representations, respectively.

After symmetry breaking operators (\ref{eop1}) and (\ref{eop2})
modify the charged and neutral currents as
\begin{eqnarray}
\label{current1}
J^{CC}_\mu=\left[1+2\alpha^\prime_j\frac{v^2}{\Lambda^2}\right]\bar
{{\nu}_j}_{L}\gamma_\mu {\ell_j}_{L}
\end{eqnarray}
\begin{eqnarray}
\label{current2} J^{NC}_\mu=\left[\frac{1}{2}+
\frac{v^2}{2\Lambda^2}(-\alpha_j+\alpha^\prime_j)\right] \bar
{{\nu}_j}_{L}\gamma_\mu {{\nu}_j}_{L}+\left[-\frac{1}{2}+
\sin^2\theta_W-\frac{v^2}{2\Lambda^2}(\alpha_j+\alpha^\prime_j)\right]
\bar {{\ell}_j}_{L} \gamma_\mu {\ell_j}_{L}
\end{eqnarray}
In this effective current subscript "L" represents the left-handed
leptons and $v$ represents the vacuum expectation value of the
scalar field and $\Lambda$ represents new physics energy scale. (For
definiteness, we take $v=246$ GeV and $\Lambda = 1$ TeV in the
calculations presented in this paper). $\alpha_j$ and
$\alpha^\prime_j$ are the coupling constants for the operators
(\ref{eop1}) and (\ref{eop2}) respectively.

One can see from (\ref{current1}) and (\ref{current2}) that the
coupling $\alpha^\prime_j$ contribute both charged and neutral
currents. But the charged current isolates it. Therefore a combine
analysis of charged and neutral currents provides us the opportunity
to discern the couplings $\alpha^\prime_j$ and $\alpha_j$.

\section{cross sections and numerical analysis}

The charged current process $\nu_\mu e^-\to \nu_e \mu^-$, called
"inverse muon decay" is described by a t-channel W exchange diagram.
As can be seen from current (\ref{current1}) that in case of
neutrino flavor universality violation $W\mu\nu_\mu$ and $We\nu_e$
vertices are modified by different operators. The cross section is
then given by the simple formula
\begin{eqnarray}
\label{cs} \frac{d\sigma(\nu_\mu e^-\to \nu_e
\mu^-)}{dy}=\frac{G_F^2}{\pi}\left(1+\frac{2v^2}{\Lambda^2}\alpha^\prime_\mu\right)^2
\left(1+\frac{2v^2}{\Lambda^2}\alpha^\prime_e\right)^2\left(2m_eE_\nu-(m_\mu^2-m_e^2)\right)
\end{eqnarray}
where $E_\nu$ is the initial neutrino energy, $m_e$ and $m_\mu$ are
the mass of the electron and muon, $G_F$ is the Fermi constant and
\begin{eqnarray}
y=\frac {E_\mu-\frac{m^2_\mu+m^2_e}{2m_e}}{E_\nu}
\end{eqnarray}
with $E_\mu$ being the final muon energy. The range of y is
\begin{eqnarray}
0\leq y\leq 1-\frac{m^2_\mu}{2 m_e E_\nu+m^2_e}
\end{eqnarray}

We studied $95\%$ C.L. bounds using two-parameter $\chi^{2}$
analysis with and without a systematic error. The $\chi^{2}$
function is given by,
\begin{eqnarray}
\chi^{2}=\left(\frac{\sigma_{SM}-\sigma_{AN}}{\sigma_{SM} \,\,
\delta_{exp}}\right)^{2}
\end{eqnarray}
where $\sigma_{AN}$ is the cross section containing new physics
effects and $\delta_{exp}=\sqrt{\delta_{stat}^2+\delta_{syst}^2}$.
$\delta_{stat}=\frac{1}{\sqrt{N}}$ is the statistical error and
$\delta_{syst}$ is the systematic error. The number of events for
inverse muon decay is proposed to be $N=7\times 10^5$ in the NuSOnG
proposal \cite{Adams:2008cm}. Therefore we assume this event number.

\begin{figure}
\includegraphics{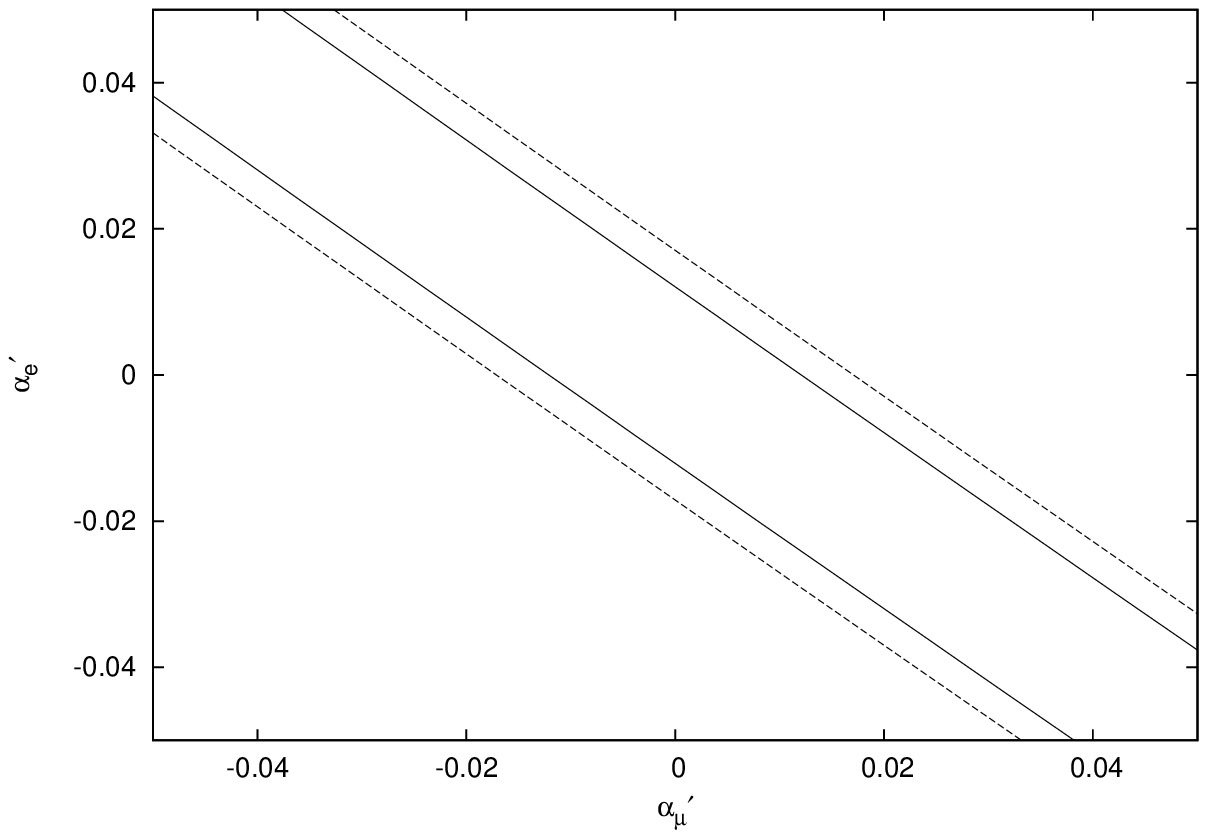}
\caption{$95\%$ C.L. sensitivity bounds on the parameter space
$\alpha^\prime_\mu - \alpha^\prime_e$.
 The area
restricted by the solid lines shows the sensitivity bound without a
systematic error and dotted lines shows the sensitivity bound with a
systematic error of the same order as the statistical one.
\label{fig1}}
\end{figure}

In Fig.\ref{fig1} we plot $95\%$ C.L. bounds on the parameter space
$\alpha^\prime_\mu - \alpha^\prime_e$ for inverse muon decay. We
consider two cases; bounds without a systematic error and with a
systematic error of the same order as the statistical one. We see
from the figure that our limit with a statistical error is about an
order of magnitude better than the limit obtained from muon decay
\cite{muondecay}.

NuSOnG experiment would provide unprecedented statistics for charged
current deep inelastic scattering from the nuclei in glass. The
expected number of events for $\nu_\mu$ charged current deep
inelastic scattering is $600\times10^6$ \cite{Adams:2008cm}. In
contrast,  NuTeV had $1.62\times10^6$ deep inelastic scattering
(NC+CC) events in neutrino mode \cite{zeller}.  Therefore NuSOnG
could provide two orders of magnitude more events. In the effective
lagrangian approach of Ref. \cite{Buchmuller:1985jz}, operators
(\ref{eop1}) and (\ref{eop2}) do not modify the quark couplings to W
and Z. Therefore the hadron tensor
remains in the standard form \cite{Blumlein:1996vs,Forte:2001ph}
\begin{eqnarray}
W_{\mu\nu}=\left(-g_{\mu\nu}+\frac{q_\mu
q_\nu}{q^2}\right)F_1(x,Q^2)+\frac{\hat{p}_\mu \hat{p}_\nu}{p\cdot
q}F_2(x,Q^2)-i\epsilon_{\mu\nu\alpha\beta} \frac{q^\alpha
p^\beta}{2p\cdot q} F_3(x,Q^2)
\end{eqnarray}
where $p_\mu$ is the nucleon momentum, $q_\mu$ is the momentum of
the gauge boson propagator, $Q^2=-q^2$, $x=\frac{Q^2}{2p\cdot q}$
and
\begin{eqnarray}
\hat{p}_\mu\equiv p_\mu-\frac{p\cdot q}{q^2}q_\mu \nonumber
\end{eqnarray}
The charged current structure functions for an isoscalar target are
defined as follows \cite{CHARM1988}
\begin{eqnarray}
\label{CC}
F_2=&&x(q_{val}+2\bar {q})+x(s-c)\nonumber\\
F_3=&&q_{val}
\end{eqnarray}
where,  $q_{val}$'s are valence quark and $q$'s are sea quark
distributions. The form factors $F_1$'s can be obtained from
(\ref{CC}) by using Callan-Gross relation $2xF_1=F_2$
\cite{Callan-Gross}. In our calculations parton distribution
functions of Martin, Roberts, Stirling and Thorne (MRST2004)
\cite{Martin:2004dh} have been used. We assume an isoscalar nucleus
$N=(p+n)/2$. This is a good assumption for a glass target as was
discussed in Ref. \cite{Balantekin:2008rc}.

New physics contributions coming from the operators in
(\ref{eop1}) and (\ref{eop2}) only modify the lepton tensor:
\begin{eqnarray}
\label{LeptonTensor}
L_{\mu\nu}=&&8\left(1+\frac{2v^2}{\Lambda^2}\,\alpha_\mu^\prime\right)^2
\left(k_\mu k_\nu^\prime+k_\mu^\prime k_\nu-k\cdot k^\prime
g_{\mu\nu}+i\epsilon_{\mu\nu\alpha\beta}k^\alpha k'^\beta \right)
\end{eqnarray}
where, $k_\mu$ and $k_\mu^\prime$ are the momenta of initial
$\nu_\mu$ and final $\mu^-$ respectively. We see from
(\ref{LeptonTensor}) that charged current deep inelastic scattering
isolates the coupling $\alpha_\mu^\prime$. It does not receive any
contribution from $\alpha_e^\prime$. As we have seen this is not the
case in inverse muon decay.

\begin{figure}
\includegraphics{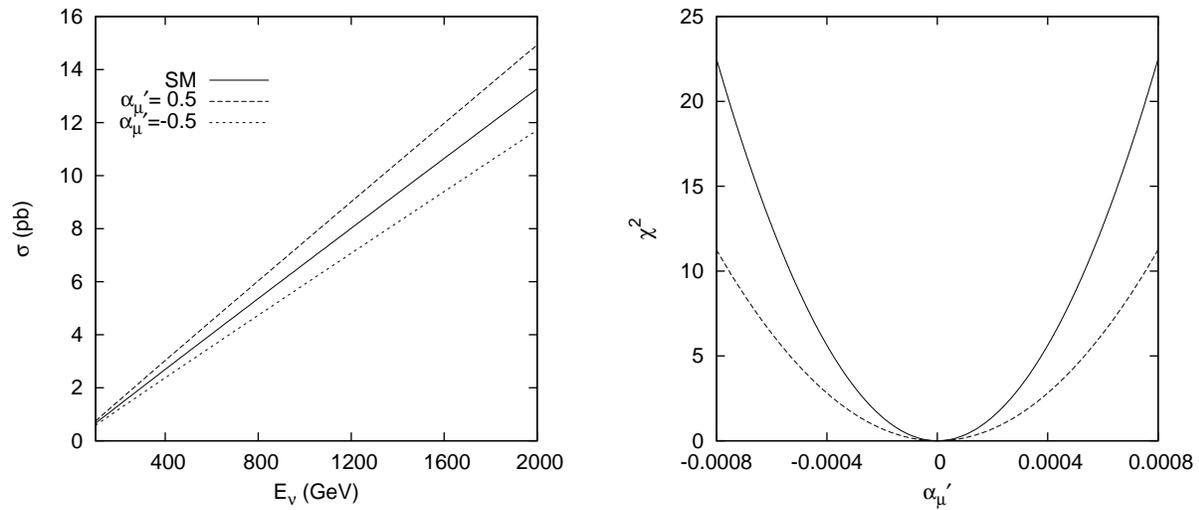}
\caption{Figure on the left shows charged current deep inelastic
scattering cross section of $\nu_\mu$ from an isoscalar nucleus as a
function of neutrino energy for the standard model (SM-solid line)
and two values of the anomalous coupling $\alpha^\prime_\mu$.
Figure on the right shows $\chi^2$ function versus
$\alpha^\prime_\mu$ with (dotted line) and without (solid line) a
systematic error. Systematic error is taken to be of the same order
as the statistical one. \label{fig2}}
\end{figure}

The behavior of the charged current deep inelastic scattering cross
section as a function of initial neutrino energy is plotted for
various values of the anomalous coupling $\alpha^\prime_\mu$ in the
left panel of Fig.\ref{fig2}. We see from this figure that deviation
of the anomalous cross sections from the SM increase as the energy
increases. Therefore high energy neutrino experiments are expected
to reach a high sensitivity to probe this anomalous coupling. In the
right panel of Fig.\ref{fig2} $\chi^2$ function versus anomalous
coupling $\alpha^\prime_\mu$ is plotted for charged current deep
inelastic scattering. $95\%$ C.L. sensitivity bound on
$\alpha^\prime_\mu$ with a systematic error of the same order as the
statistical one is $\mid\alpha^\prime_\mu\mid \leq 5\times10^{-4}$.
This bound is two order of magnitude better than the LEP bound
obtained from $W^+\to\mu^+\nu_\mu$  decay  \cite{Yao:2006px}.

\begin{figure}
\includegraphics{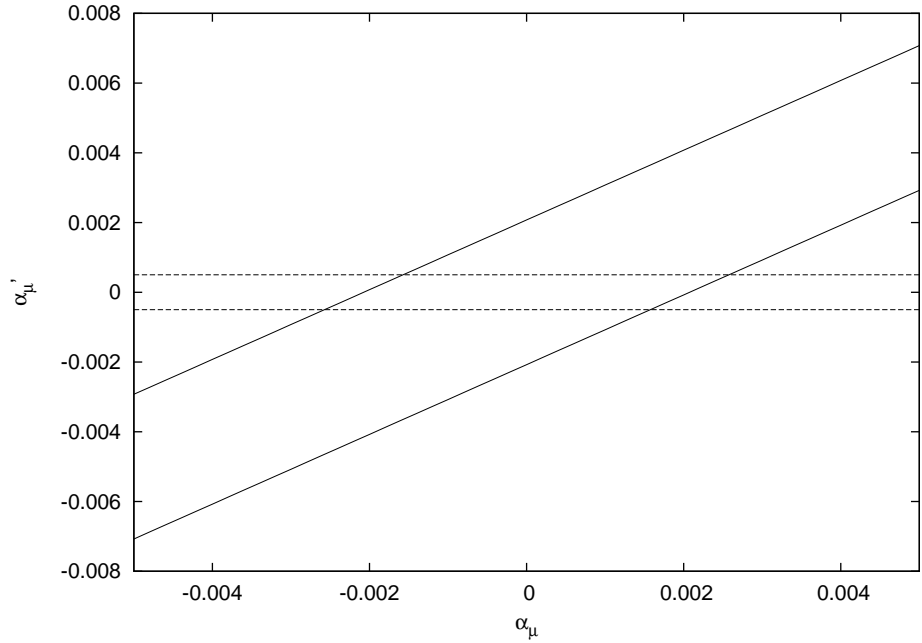}
\caption{Neutral current deep inelastic scattering limits (solid
lines) of Ref.\cite{Balantekin:2008rc} and charged current deep
inelastic scattering limits (dotted lines) on $\alpha^\prime_\mu$.
Limits are at 95\% C.L. with a systematic error of the same order as
the statistical one. \label{fig3}}
\end{figure}

Neutral current deep inelastic scattering was analyzed in
Ref.\cite{Balantekin:2008rc} in detail. We see from (\ref{current2})
that it receives contributions both from couplings
$\alpha^\prime_\mu$ and $\alpha_\mu$. Therefore it is impossible to
set a limit on $\alpha_\mu$ independent from $\alpha^\prime_\mu$
with neutral current deep inelastic scattering alone. It is rather
possible to set a limit on $\alpha^\prime_\mu - \alpha_\mu$ plane.
On the other hand charged current deep inelastic scattering isolates
the coupling $\alpha^\prime_\mu$ and therefore combining charged and
neutral current scattering we can extract the limit on $\alpha_\mu$.
In Fig.\ref{fig3} we plot neutral current deep inelastic scattering
limits (solid lines) of Ref.\cite{Balantekin:2008rc} and charged
current deep inelastic scattering limits (dotted lines) on
$\alpha^\prime_\mu$. Intersection of these two bounds gives us the
limit $\mid\alpha_\mu\mid\leq 2.6\times10^{-3}$ at 95\% C.L..

Universality assumption of the standard model that states $\nu_e$,
$\nu_\mu$ and $\nu_\tau$ couple with the same strength to W and Z at
the tree level has been discussed in Ref.\cite{Balantekin:2008rc}
and the NuSOnG bounds on universality violation have been obtained
from neutral current reactions. Authors obtained the limits
$\mid\alpha^\prime_e-\alpha^\prime_\mu\mid \leq
0.074\,(\alpha_\mu=0)$ and $\mid\alpha_e-\alpha_\mu\mid \leq
0.071\,(\alpha^\prime_\mu=0)$ under the assumption that only one
type of operator $O_\mu$ or $O^\prime_\mu$ contributes to the
effective lagrangian. This assumption is necessary since neutral
current receives contributions both from $O_\mu$ and $O^\prime_\mu$.
On the other hand we see that the couplings $\alpha^\prime_\mu$ and
$\alpha_\mu$ can be constrained separately by studying neutral and
charged current reactions together. By combining charged current
deep inelastic scattering limit with the results of
Ref.\cite{Balantekin:2008rc} we obtain $\mid\alpha_e-\alpha_\mu\mid
\leq 0.071$.  Similarly combining the limit
$\mid\alpha^\prime_\mu\mid \leq 5\times10^{-4}$ with the bound in
Fig.\ref{fig1} we obtain $\mid\alpha^\prime_e-\alpha^\prime_\mu\mid
\leq 0.019$. These limits are at 95\% C.L. with a systematic error
of the same order as the statistical one. We consider both of the
operators (\ref{eop1}) and (\ref{eop2}) and do not assume that only
one of them contribute to the effective lagrangian at once.

\section{Conclusions}
In this paper we investigated signatures for deviation from standard
model predictions in neutrino-W boson couplings via charged current
reactions in the NuSOnG. We showed that charged current deep
inelastic scattering can explore new physics signatures with a
sensitivity of order $10^{-4}$. Therefore NuSOnG provides
exceptional prospect to probe neutrino-W couplings. We also explored
the NuSOnG potential to probe neutrino flavor universality violation
in neutrino-W and neutrino-Z couplings. Combining the results for
charged current reactions with the results for neutral current
reactions of Ref.\cite{Balantekin:2008rc}  we achieved a complete
analysis. We deduce that NuSOnG has a great potential to probe
possible universality violation in neutrino-W and neutrino-Z
couplings.

\begin{acknowledgments}
We thank J. Conrad for discussions. This work was supported in part
by the U.S. National Science Foundation Grant No.\ PHY-0555231 and
in part by the University of Wisconsin Research Committee with funds
granted by the Wisconsin Alumni Research Foundation. \.{I}.
\c{S}ahin acknowledge support through the Scientific and Technical
Research Council (TUBITAK) BIDEB-2219 grant.
\end{acknowledgments}

\pagebreak

\pagebreak

\end{document}